# Using Programmable Graphene Channels as Weights in Spin-Diffusive Neuromorphic Computing

Jiaxi Hu, Gordon Stecklein, Yoska Anugrah, Paul A. Crowell, and Steven J. Koester, *Fellow, IEEE*

*Abstract*— A graphene-based spin-diffusive (GrSD) neural network is presented in this work that takes advantage of the locally tunable spin transport of graphene and the non-volatility of nanomagnets. By using electrostatically gated graphene as spintronic synapses, a weighted summation operation can be performed in the spin domain while the weights can be programmed using circuits in the charge domain. Four-component spin/charge circuit simulations coupled to magnetic dynamics are used to show the feasibility of the neuron-synapse functionality and quantify the analog weighting capability of the graphene under different spin relaxation mechanisms. By realizing transistor-free weight implementation, the graphene spin-diffusive neural network reduces the energy consumption to 0.08-0.32 fJ per cell·synapse and achieves significantly better scalability to its digital counterparts, particularly as the number and bit accuracy of the synapses increases.

## I. INTRODUCTION

Neuromorphic systems have recently attracted a great deal of research interest, in both software and hardware implementations, for non-von-Neumann computing applications [1],[2]. Instead of being controlled by a centralized core, neuromorphic computing is characterized by its massive parallelism, which is enabled by an enormous number of simple units, called "neurons" [3]. The communication between neurons occurs through "synapses," where a "weight" describes the signal strength. In general, the behavior of one neuron can be written as [4], [5]:

$$y_j = f(\textstyle\sum_{i=1}^{N} w_{ji} x_i + b) \ , \tag{1}$$

where $y_j$ is the output of the present neuron, $x_i$ is the output from its $N$ neighbors and $b$ is the biasing condition. The function $f(x)$, also called the "activation function," is a nonlinear transfer function which describes the relationship between state and its output. $w_{i,j}$ is the weight that describes the strength of this connection in which the number, precision, and complexity of these connections determines the functionality and performance of neuromorphic architectures.

Because neuromorphic systems merge the computation and memory functionality, devices with non-volatile properties have been intensively explored. Cross-bar architectures [6], [7] using phase change (PC) devices [8], [9] and magnetic tunnel junctions (MTJs) [10], [11] have been studied as dense synapse arrays where the weights are programmed as the charge resistance along the current path. However, as the summation operation occurs in the charge current domain, extra transistor components are often required as independent current sources, and these consume additional space and power. In addition, compared to the digital weights often used in PC- and MTJ-based architectures, analog weights are attractive due to their compact size and more precise tunability [12], [13].

Spin current, which is the net flow of non-equilibrium spins, provides an alternative path to realize current-mode weight summation. Combining the lateral spin valve (LSV) geometry [14] and spin-transfer-torque switching (STT) [15], [16], spin-diffusive interconnects have been proposed [17], [18]. The initial exploration of spin-diffusive communication also shows promising results in Non-Boolean type computations, such as cellular neural networks (CNNs) [19]. However, because of the limitation of spin diffusion length, $\lambda_s$, of conventional non-magnetic (NM) spin channels, the energy dissipation of spin-diffusive communication increases significantly with the transport distance [20].

Graphene, because of its weak intrinsic spin-orbit coupling (SOC), has a long spin lifetime, $\tau_s$, and long $\lambda_s$, which allows it to act as an efficient spin-diffusive interconnect. At room temperature, experimental values of $\tau_s$ [21] and $\lambda_s$ [22], [23] are at least 2 orders of magnitude better than metallic channels. As a result, at scaled dimensions (sub µm), a significant portion of the non-equilibrium spins is able to reach the output, where contact-induced spin relaxation [24] must be taken into account, an effect which is often ignored in the presence of high-resistance or tunnel barrier contacts. In graphene spin channels, two kinds of very distinct ferromagnet/graphene (FM/Gr) interfaces can be formed: tunneling and transparent interfaces. They can be used to realize the non-reciprocity of information

This manuscript is submitted on November, 30th, 2017 for review. This work is sponsored by C-SPIN, a funded center of STARnet, a Semiconductor Research Corporation (SRC) program sponsored by MARCO and DARPA. J.H., Y.A. and S.J.K. are with the Department of Electrical and Computer Engineering, and G.S. and P.A.C. are with the School of Physics and Astronomy at the University of Minnesota, Minneapolis, MN 55455 USA (Contact e-mail: skoester@umn.edu)

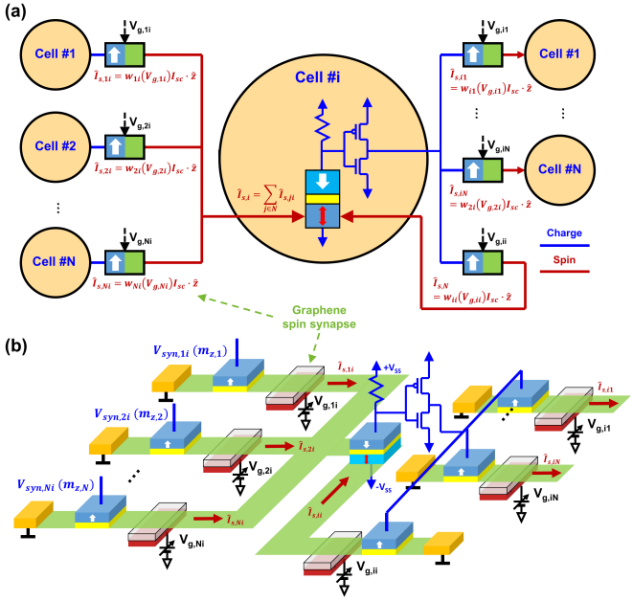

Fig.1 (a) Schematic diagram and (b) vision concept of GrSD CNN design. Synapse-to-neuron communication is in the spin current domain (red), which is controlled by the local electrostatic gating (black) while neuron-to-synapse communication uses charge voltage signals (blue). The weight programming circuits are electrically decoupled from the computation network.

flow [25], which is necessary for current-mode computation [26]. More importantly, graphene has very high mobility and a low density of states (DOS) that enables tunable spin transport [27], [28], which is not available in metallic spin channels. Experimentally, Han *et al.* [29] and Cho *et al.* [30] have shown that $\lambda_s$ in graphene can be modulated by changing the carrier concentration using electrostatic gating.

In this paper, we present a graphene-based synapse concept that allows neural networks to perform analog weighted input summation using spin diffusion current, and perform a nonlinear transfer function in the charge voltage domain. We show that using local electrostatic gating and asymmetric FM/graphene interface, scalable and energy efficient gate-controlled spin-current sources can be realized to support the spin-current mode calculation. As shown in Fig. 1, the computation network and the weight programming circuits are electrically decoupled. A low-pass filtering CNN is taken as an example case to evaluate this concept, and the functionality of network is simulated under constant and pulsed supply voltages where the trade-off between computation accuracy and energy consumption can be utilized in designing more energy efficient networks. In neuromorphic applications where a large number of accurate connections are needed between neurons, the transistor-free graphene spin synapse design is more scalable compared to its counterparts realized with digital weights.

## II. TUNABLE SPIN DIFFUSION IN GRAPHENE

Graphene, as a hexagonally arranged two-dimensional sheet of carbon atoms, has a zero-gap band structure [31], which gives it a linear energy dispersion near its Dirac point [32].

$$E_F = \pm \hbar v_F |k|, \tag{2}$$

where $k$ is the electron/hole wavenumber, $E_F$ is the energy level relative to Dirac point, $\hbar$ is the reduced Planck constant and $v_F$ is the Fermi velocity in graphene. As a result, graphene has a linear DOS:

$$g_{C,V}(E_F) = \frac{g_s g_v}{2\pi} \frac{|E_F|}{(\hbar v_F)^2}, \tag{3}$$

where $g_C$ and $g_V$ are the conduction and valence band DOS, and $g_s = g_v = 2$ are the spin and valley degeneracy, respectively. The carrier distribution in graphene follows the Fermi-Dirac distribution. In the limit of $E_F \gg k_B T$, where $k_B$ is Boltzmann's constant and $T$ is temperature, the graphene carrier density is proportional to $|E_F|^2$ (Fig. 2a). By using electro-static gating, the carrier concentration, $|n - p| = C_{tot} V_g$, can then be tuned as shown in Fig. 2, where $n$ and $p$ refer to the electron and hole concentration, $V_g$ is the applied gate voltage, and $C_{tot}$ is the effective gate capacitance taking into account the series combination of dielectric and quantum capacitances [33].

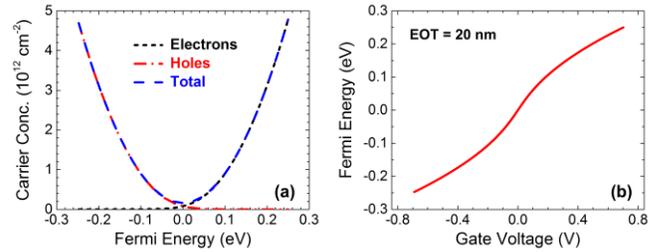

Fig. 2. (a) Carrier concentration dependence on relative Fermi energy in single layer graphene. (b) Fermi energy dependence on gate voltage, $V_g$, assuming an equivalent oxide thickness (EOT) of 20 nm, and a Dirac point at $V_g = 0$.

Experimentally, the spin relaxation in graphene is found to come from two distinct mechanisms: the Elliott-Yafet (EY) mechanism, which occurs during momentum scattering, and the D'yakonov-Perel (DP) mechanism, which arises from dephasing due to spin precession about randomly fluctuating spin-orbital fields [27]. The spin diffusion lengths governed by the EY and DP mechanisms, $\lambda_{S-EY}$ and $\lambda_{S-DP}$, respectively, can be described as

$$\lambda_{s-EY} = \frac{\pi \hbar^2 v_F}{\sqrt{2} e^2 \Delta_{EY} R_{sq}} \propto |E_F|^2, \tag{4a}$$

$$\lambda_{s-DP} = \frac{\hbar v_F}{2\sqrt{2} \Delta_{DP}}, \tag{4b}$$

where $\Delta_{EY}$ and $\Delta_{DP}$ describe the EY and DP scattering strength, respectively, and $R_{sq}$ is the sheet resistance of the graphene channel. Based upon (4a) and (4b), the theoretical dependence of $\lambda_{S-EY}$ and $\lambda_{S-DP}$ on $E_F$ are shown in Fig. 3. As the figure shows, the $E_F$-dependence of $\lambda_{S-EY}$ is parabolic, due to the inverse dependence of $R_{sq}$ on $E_F^2$. On the other hand, in the DP mechanism, the diffusion constant and spin lifetime have opposite gate dependences, leading to a relatively constant spin diffusion length. Previous work showed that DP-relaxation dominated graphene can realize a spin-demultiplexing function [34]. However, for neural networks, this paper points out that EY-relaxation dominated single-layer graphene can create highly tunable spin synapses, which will be discussed in the following section.

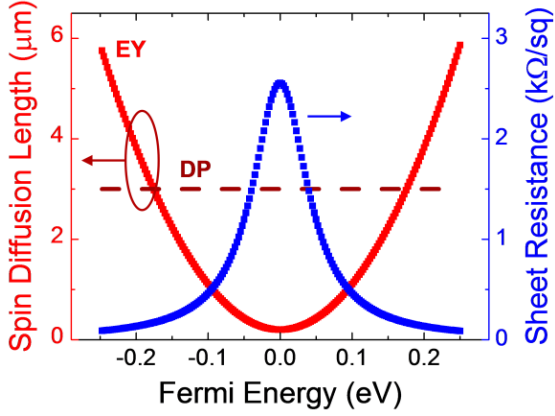

Fig. 3. Spin diffusion length modulation from EY (red) and DP (dark red) spin relaxation mechanism, and sheet resistance (blue) in single layer graphene. The calculations assume $\Delta_{EY} = 12$ meV and $\Delta_{DP} = 77$ μeV.

## III. GRAPHENE SPIN SYNAPSES

### A. Spin Current Modulation in Graphene Channels

The design of the spin-diffusive graphene synapse is based on the LSV geometry (Fig. 1b). Here, each synapse operates as a voltage controlled spin-current source (Fig. 4), where the synapse takes a charge voltage, $V_{syn}$, as its input and transfers it into the output spin diffusion current $I_{s,out}$ which is absorbed by the detector magnets. Fig. 5a shows the gate modulation of the spin current transmission coefficient $T_s = |\hat{I}_{s,out}|/|\hat{I}_{s,in}|$, in a pure graphene interconnect, under EY and DP dominant spin relaxation, respectively. Based on the relative resistance value of the π-network module [35], the analytical expression of $T_s$ can be written as:

$$T_s = \frac{R_{sf\pi}}{R_{se\pi} + R_{sf\pi}} \cdot \frac{R_{sf\pi}}{R_{s,det} + R_{sf\pi}}, \qquad (5)$$

where $R_{se\pi}$ and $R_{sf\pi}$ refer to the series and shunt spin resistance in the π -module network which are proportional to $R_{sq}$. In the case of a transparent FM/graphene interface [36] (the detector resistance, $R_{s,det}$, is negligible comparing to $R_{sf\pi}$) the synapse essentially acts as a spin current divider between $R_{sf\pi}$ and $R_{se\pi}$.

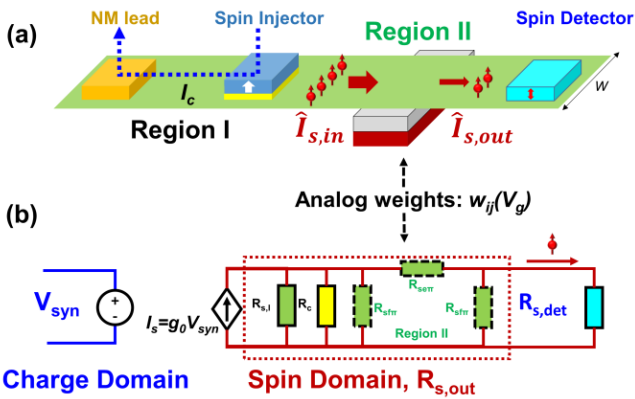

Fig. 4. (a) Schematic and (b) equivalent circuit model of graphene spin-diffusive synapse as a charge tunable voltage-controlled spin current source.

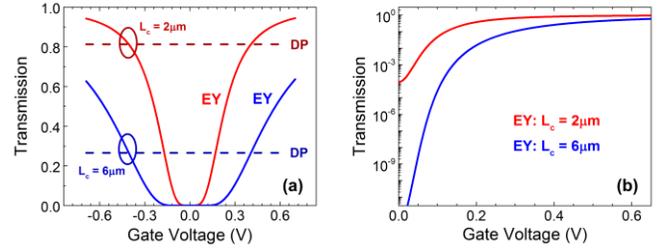

Fig. 5. (a) Gate dependence of spin transmission efficiency, $T_s$, on a linear scale for a single layer graphene spin-interconnect with EY or DP as the dominant spin relaxation mechanism (b) Log scale of $T_s$ with EY spin relaxation only. The plots assume zero spin resistance at the detector magnet.

With EY spin relaxation, the dynamic range of $T_s$ can be made very large and tends to increase with the graphene synapse length, $L_c$. Assuming $\Delta_{EY} = 12$ meV, and $L_c = 2$ μm, for a local backgate voltage, $V_g$, varied between 0 and $\pm 0.7$ V, $\lambda_{s-EY}$ will swing from 0.2 to 6.0 μm, corresponding ~$10^4$ modulation of $T_s$, as shown in Fig. 5b. However, in the case of DP relaxation, $T_s$ is independent of gate voltage.

To perform current-mode neuromorphic computing, the current sources must be able to provide large enough output resistance to support high fan-in structures. As shown in Fig. 4, in the limit of $L_c \ll \lambda_s$, the output resistance of a graphene spin synapse mainly determined by the right shunt spin resistor: $R_{s,out} = R_{sf\pi} \approx \frac{R_{sq}\lambda_s}{W}$, where $W$ is the synapse width. Conversely, in the limit of $L_c \gg \lambda_s$, the output resistance of graphene spin synapse is determined by the parallel combination of the injector contact resistance and the graphene spin resistance between injector and ground: $R_{s,out} = R_c || \frac{R_{sq}\lambda_s}{W}$. Since a tunnel barrier is introduced at injector to promote spin injection efficiency [37], $R_c$ would be much higher than $\frac{R_{sq}\lambda_s}{W}$, and therefore, in either case discussed above, the output spin resistances, converges to $\frac{R_{sq}\lambda_s}{W}$.

### B. Cell Design

In the proposed GrSD CNN shown in Fig. 1a, each cell is composed of an MTJ, a reference resistor and a dual-rail inverter [38]. The state of each cell is represented by the free layer (FL) magnetization of the MTJ, which is a non-volatile state variable. Therefore, the computation state of the network can be well preserved when the power is off. To reduce the switching energy, materials with perpendicular magnetic anisotropy (PMA), low saturation magnetization $M_s$ and damping factor $\alpha$ are preferred [39]. Table 1 summarizes the material parameters used in this work.

The resistor and MTJ form a voltage divider between $\pm V_{sense}$ that provides an output swing between $\pm 0.2V$ by setting $R_{ref} = \sqrt{R_H \cdot R_L}$, where $R_H$ and $R_L$ refer to the high and low resistance state of the MTJ, respectively [40]. To reduce the read disturbance, an MTJ with high resistance and large tunneling magnetoresistance ratio is preferred. Through the inverter, $V_{out}$ is amplified to $\pm V_{dd}$, which will be the voltage applied at the corresponding synapses:

$$V_{syn} = |V_{dd}| \cdot f(m_z) \approx |V_{dd}| \cdot \frac{|m_z + 0.2| - |m_z - 0.2|}{0.4}, \qquad (6)$$

| Name | Value | Name | Value | Name | Value |
|---|---|---|---|---|---|
| Transistor gate capacitance, $C_g$ | 0.1 fF | Magnetic material [40] | Mn₃Ga | Graphene channel width | 20 nm |
| 15-nm-node transistor on-resistance, $R_{ON}$ | 4 kΩ | Spin injection polarization | 90% | Graphene spin channel oxide thickness | 48 nm |
| Transparent FM/graphene contact resistance [37] | 10 Ω·μm | Damping constant | 0.003 | Graphene spin channel length | 2 μm |
| Tunneling FM/graphene contact resistance | 100 Ω·μm | Saturation magnetization | 0.11 MA/m | Spin MTJ resistances, $R_{high}/R_{low}$ | 250 / 25 kΩ |
| Transistor supply voltage, $V_{dd}$ | 0.5 V | Magnetic anisotropy | 0.21 MJ/m³ | Reference resistor, $R_{ref}$ | 80 kΩ |
| Sensing voltage, $V_{sense}$ | 0.5 V | Nanomagnet size | 40×10×3 nm³ | Inverter charging delay | 80 ps |
| Transistor threshold voltage, $V_{th}$ | 0.2 V | # of synapses per cell | 4 + 1 | Surge current at each synapse | 100 μA |

Table 1. Material and device parameters used in GrSD CNN simulation.

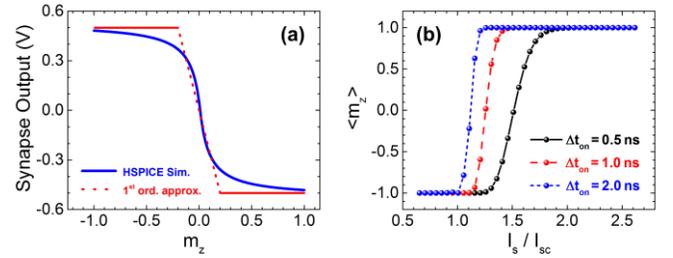

Fig. 6. (a) Output charge voltage as a function of input magnetization state simulated by HSPICE (blue) and its linear approximation (red). The linear function is used in the CNN simulations in the next section. (b) Averaged z-magnetization $\langle m_z \rangle$ of magnet FL based upon 100 Monte Carlo simulations under different spin current pulse width.

where $m_z$ is the $z$-component of free layer magnetization. The relation in (6) is the 1$^{st}$ order approximation used in an HSPICE simulation using the 16-nm technology node transistor model (Fig. 6a) [41].

### C. Transfer Function

In this CNN, the transfer function takes spin current as its input and provides an output voltage at its synapses, $V_{syn} = f(\hat{I}_s)$. The magnetization of the FL magnet is switched by STT, where the dynamics are described by stochastic Landau-Lifshitz-Gilbert equation (s-LLG), that includes the magnetic anisotropy field, demagnetizing field and random thermal fluctuations [42].

Since most of the spin-torque induced switching finishes in the first few nanoseconds, employing pulsed operation would save the energy consumption on the spurious charge current path [43]. By pulsing the dual-rail voltage supply for the inverter in each cell, the output spin current is also transient, and the switching behavior of the PMA magnet is described by its probability density function. As shown in Fig. 6b, with a longer input pulse duration, the switching of the free layer magnetization becomes sharper, which means the switching tends to be more deterministic. However, with a narrower pulse, depending upon the initial angles of magnetization [42], some magnets may not be able to switch in time. It indicates that the transfer function from the spin current to the charge voltage domain (Figs. 6) has a tunable slope, which could be used for the optimization between network energy consumption and performance [44].

### D. Weight Programming

Asymmetric interfaces for spin injection/detection can promote the directionality of spin information flow [25]. At the injector, a tunneling interface is preferred to match the spin resistance between the FMs and graphene, while at the detector, a transparent interface promotes spin current absorption and STT switching. The weight parameter describes the spin current intensity that diffuses from the synapses to the neuron bodies. The weight is defined as the output spin current normalized to the critical current of the FL PMA magnets, $|I_{SC}|$, when $V_{syn}$ saturates to $\pm V_{dd}$, and can be expressed as:

$$w_{ji} = syn(m_{z,inj}) T_s (V_{s,ji}) \cdot \frac{P \cdot |V_{dd}|}{|I_{SC}| \cdot (R_C + R_{Gr,GND})} , \quad (7)$$

where the magnetization direction of the spin injector, $m_{z,inj}$, determines the sign of the weight; $R_C$ and $R_{Gr,GND}$ are the charge resistance of the tunnel contact and graphene resistance in region I, respectively; $P$ is the spin injection polarization, $V_{dd}$ is the supply voltage. A unit weight, $|w_{ij}| = 1$, corresponds to the case that the synapse outputs have the same magnitude as the $|I_{SC}|$ of the FL PMA magnets: $|I_{s,out}(i \to j)| = |I_{SC}|$. The critical switching spin current of the FL PMA magnets is estimated as [42]:

$$|I_{SC}| = 2e\alpha\Delta k_B T / \hbar P , \quad (8)$$

where $\alpha$ is the damping constant, $e$ is the electron charge, and $\Delta$ is the thermal stability factor. Fig. 7 shows the gate dependence of the applied weight at a 40-nm-wide synapse. In the case of EY spin relaxation, the applied weight, $w$, can be modulated from nearly zero to 25 with a channel length of 2 μm. Unlike the case discussed in Fig. 5, with finite $R_C$, $R_{s,I}$ and non-zero $R_{s,det}$, DP spin relaxation is also able to tune its weight by a factor of ~3 to 5, which is a result of contact-induced modulation [24]: the relative ratio between graphene and contact spin resistance changes with $R_{sq}$. Because $\lambda_{DP}$ has no gate voltage dependence, fundamentally, the modulation of DP spin-relaxation is limited to the charge resistivity modulation of graphene, which is usually less than a factor of 10 at room temperature.

On the other hand, with EY-type of spin relaxation, the weight value can be tuned exponentially: $|w_{ij}|^* \propto e^{-L_C/\lambda_s(V_g)}$, where $L_C$ is the distance between the spin injectors and neuron body. As shown in Fig. 7c, this type of modulation can provide 2 to 9 orders of magnitude dynamic range for the weights. However, the strong modulation of EY spin-relaxation comes with a limited absolute weight value. This problem can be solved by increasing the supply voltage $V_{dd}$ or introducing negative a self-feedback channel as shown in Fig. 1b. The negative self-bias reduces the thermal stability of the FL magnet only when the network is active (powered on), yet preserves the cell non-volatility when the power is off.

### E. Analog Weighted Summation

The spin current from different synapses is summed up in cell $i$ where the graphene channel merges according to:

$$\hat{I}_{s,NeuronBi} = I_{SC} \cdot \left( \sum_{j=1}^{N} w_{ji}(V_{s,ji}) \cdot m_{j,z} \right), \quad (9)$$

where $m_{j,z}$ is the $z$-component of the free layer magnetization in cell $j$. Therefore, the CNN templates can be programmed and operated in the spin-current domain [45] according to

$$\hat{I}_{s,j} = \left( \sum_{j \in N,N} A_{ji} y_j + \sum_{j \in N,N} B_{ji} u_j + i_{bias} \right), \quad (10)$$

where $y_j = I_{SC} \cdot f(m_{z,j})$ and $u_j$ represents the output and input of each cell. $i_{bias}$ can be set as an external constant spin current input, while $i$ and $j$ refer to the index of receiving and transmitting cells.

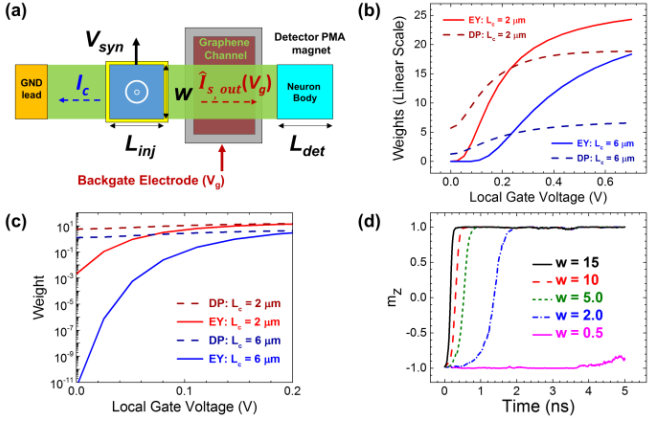

Fig. 7. (a) Neuron connected with a single graphene spin synapse (top view), the local gate dependence of the applied weights under (b) linear and (c) log scale. (d) Switching dynamics of FL PMA magnet under different applied weight value.

Fig. 8 shows the simulation of a neuron (Fig. 8a) with two input synapses under different bias combinations, $(V_{syn1}, V_{syn2})$, that operate the device as an analog spin adder/subtractor. Fig. 8b shows the switching dynamics simulated by a 4-component charge/spin-circuit model [35]. The corresponding computation table is plotted in Fig. 8c. The table shows that the summed value of weights, $w_1+w_2$, have the same switching delay as the case with single-input devices (Fig. 7d). This indicates that the cross spin diffusion current between two input synapses is negligible so that $w_{tot}(V_{g1}, V_{g2}) = w_1(V_{g1}) + w_2(V_{g2})$ is valid. This behavior is a direct result of graphene's high spin resistance.

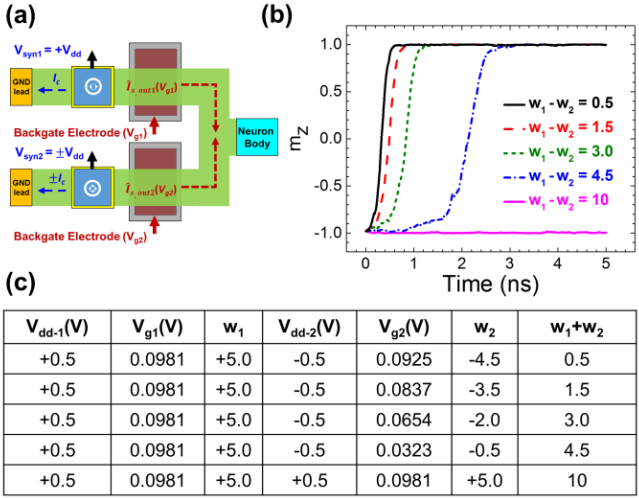

Fig. 8. (a) Schematic of cell with two input synapses, (b) Simulation of the corresponding switching dynamics (c) Weight programming table of the spin adder/subtractor.

## IV. SIMULATION OF CNN NETWORK

The functionality and the energy-delay performance of the graphene spin synapses in larger scale CNN have been estimated using a low-pass filtering application as an example. Fig. 9 shows the simulation of a 26×24 pixel image of a hand-drawn character "a" with 15% initial noise. The input binary image is first programmed as the magnetization of FL magnets. The STT-induced switching is simulated using the framework described previously. The template used in the low-pass filter application is [45]:

$$A = W_0 \times \begin{bmatrix} 0 & 1 & 0 \\ 1 & 1 & 1 \\ 0 & 1 & 0 \end{bmatrix}, B = 0, I_{bias} = 0 \quad , \quad (11)$$

where $W_0$ is the weight value that determines the speed and power consumption of the CNN system. The computation results are compared to the reference solution obtained using the standard RC CNN dynamical equations [46].

Fig. 9 describes the filtering process under four different situations: in the first two cases, the supply voltage is constant, and synapses are programmed with different $W_0$ ($W_0 = 1$ and 2 for the top and second row respectively). In the other two cases, the CNN is operated using a pulsed supply voltage with 1 or 2 ns pulse width. After calculating the two-dimensional Fourier transform of the image, we integrate its power spectrum starting from spatial frequency $\sqrt{f_x^2 + f_y^2} = 1/3$ to represent the power of "high-spatial-frequency-components" (hSFc) which is in turn used as a quantitative variable to track the image evolution during the filtering process. Under static bias, with $W_0 = 1$, the percent of hSFc reduces from 20.3% to 6.1 %, which indicates that the steady state solution has been reached after about 4 ns. By increasing the value of $W_0$ from 1 to 2, the convergence of the filtering process becomes faster and reaches a lower ultimate hSFc level of 4.8%.

The dynamics of noise filtering process can be decomposed into the switching of 3 basic patterns shown in Fig. 10. Based on the value of template $A$ (11), the middle pixel absorbs the spin current injected from its four nearest neighbors and itself. The summed spin current is then equal to: $3W_0 \cdot |I_{SC}|$, $W_0 \cdot |I_{SC}|$ and $-W_0 \cdot |I_{SC}|$ for Figs. 10a, b and c, respectively. The corresponding switching delays can also be estimated using

$$\langle \Delta t \rangle = \tau_0 \times \frac{I_{SC}}{I_S - I_{SC}}, \quad (12)$$

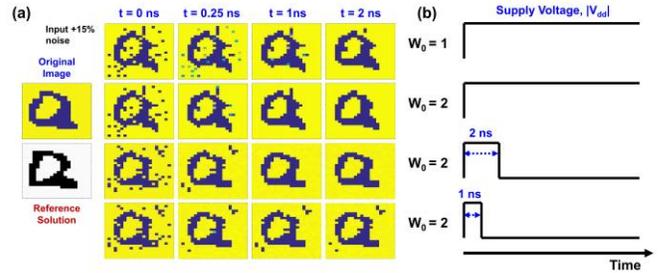

Fig. 9. (a) Simulation of a low-pass filtering application on a hand-drawn character "a" (30×30 image) with 15% initial noise. Yellow and blue pixels indicate the $m_z$ along the ±z direction. (b) The supply voltage pulse shapes used in the CNN simulation.

| $V_{dd-1}(V)$ | $V_{g1}(V)$ | $w_1$ | $V_{dd-2}(V)$ | $V_{g2}(V)$ | $w_2$ | $w_1+w_2$ |
|---|---|---|---|---|---|---|
| +0.5 | 0.0981 | +5.0 | -0.5 | 0.0925 | -4.5 | 0.5 |
| +0.5 | 0.0981 | +5.0 | -0.5 | 0.0837 | -3.5 | 1.5 |
| +0.5 | 0.0981 | +5.0 | -0.5 | 0.0654 | -2.0 | 3.0 |
| +0.5 | 0.0981 | +5.0 | -0.5 | 0.0323 | -0.5 | 4.5 |
| +0.5 | 0.0981 | +5.0 | +5.0 | 0.0 | +5.0 | 10 |

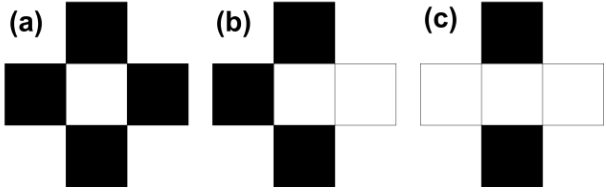

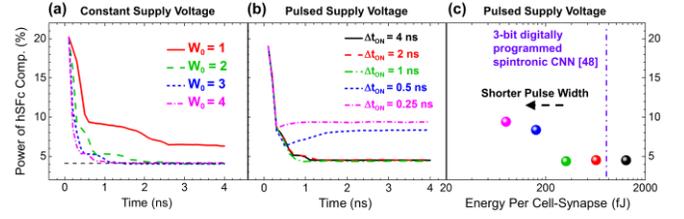

Fig. 10. The three basic switching patterns that are involved in the low-pass filtering applications using the template in (13). Black and white panel correspond to neurons with magnetization along $+z$ and $-z$ direction.

Fig. 11. (a) The percent of hSFcs during the noise filtering based on s-LLG simulation under a static power supply but with different $W_0$ in template $A$. (b) s-LLG simulated image convergence under pulsed supply voltages. The clock cycle is 8ns and the pulse width varies from 2 ns to 0.25 ns. (c) The energy consumption of the pulsed CNN plotted in (b).

where $\tau_0$ is the time scale of Sun's model [42]. In Fig. 10c, since the center pixel has no net spin current from its neighbors, the switching can only happen due to random thermal fluctuations. Under the same bias condition, pattern (a) switches on average $\frac{3W_0-1}{W_0-1}$ times faster than pattern (b).

As discussed previously, by controlling the pulse width of the supply voltage, it is possible to trade off computational accuracy for reduced energy consumption. Fig. 9 shows that the first two simulations perform better, but come with the price of extra leakage power (in the second row, after 1 ns, most of the cells are not switching). As shown in the 3rd and 4th row of Fig. 9, the residual hSFc in pulsed operations increases with shorter pulse width. It can be explained by the fact that the pattern in Fig. 10b needs a longer time for switching. As the net input spin current approaches its critical switching current, the estimation of switching delay needs to use more accurate models rather than Eq. (12) to include the stochastic switching events. Under 1-ns pulse width, some of the pattern in Fig. 10b may not be able to switch in time so that, in the final solution, there is more residual hSFc. It can also be understood as that the slope of $\langle m_z \rangle$ vs $|\hat{I}_s|$ becomes softer with a shorter pulse width (Fig. 6b), this allows more hSFcs passes through the filter. If the thermal stability of the FL magnets is reduced, the slope of $\langle m_z \rangle$ vs $|\hat{I}_s|$ will become even softer and more pulse-width dependent, indicating the switching events become more non-deterministic.

The energy consumption (per cell) of the graphene spin-CNN is calculated as:

$$E = \frac{|V_{dd}|^2}{R_C+R_{Gr,I}} \cdot N_{syn} \cdot \Delta t_{ON} + E_{read} + E_{inverter},\qquad(13)$$

where $N_{syn} = 5$ represents the 5 connections in each cell, including the communication between 4 nearest neighbors and one self-feedback path (11). $E_{read}$ and $E_{inverter}$ are the leakage energy of the voltage divider and dynamical charging/discharging energy of the inverter, respectively. For the 1-ns pulse width, the energy consumption is 0.32 fJ per cell·synapse, which is 4 times lower than the digital spintronic CNNs analyzed in [45], though this is mainly because of the use of PMA Heusler alloy as the FM material of the FL magnets, in which the low damping and low saturation magnetization helps reduce the energy consumption of STT-switching. As shown in Fig. 11, most of the switching events are able to finish in the first 1-2 ns, but by trading off computation accuracy ($\Delta t_{ON} < 1 ns$), the energy consumption of graphene spin-diffusion CNNs can be reduced to 80aJ per cell·synapse.

Another key advantage of the graphene-based spin synapses is their capability for analog programmed weights [47] in designing sophisticated CNN applications where more synapse connectivity and higher weight accuracy are necessary [48]. Comparing to digitally programmed spintronic weights, the gate dependence of EY-spin relaxation allows graphene spin synapses to realize high weight resolution in a more compact dimension. Assuming the applied gate voltage has a 1 mV resolution, the weight resolution is defined as:

$$\Delta w = \left|\frac{dw}{dV_g}\right|_{max} \times \Delta V_g,\qquad(14)$$

$$NoB = \log_2\left|\frac{w_{max}}{\Delta w}\right|,\qquad(15)$$

where $NoB$ is "number of binary bits" required to achieve the same weight resolution. As shown in (14) and (15), $NoB$ is determined by the maximum achievable weights, $|w_{max}|$, and the maximum slope of weight modulation, $dw/dV_g$. As shown in Fig. 12, when $L_c = 2\,\mu m$ (6 $\mu m$) the graphene channel length has $NoB = 8$ (9). For digital weights, these bit resolutions could lead to significant increases in weight-reprogramming energy and footprint area for the CNN. Based on the previous benchmark results on the recall accuracy in associative memory applications, increasing from 3-bit to 8-bit weight resolution, increases the required synapse count from ~20 to ~50 [19].

In the digital programmed spintronic CNNs proposed by Pan et al., each binary level of weight requires two transistors and one spin injection component. For an 8-bit weight value, 16 driving transistors in each synapse are required. Based on the applied current level, the transistor channel width varies from $W_{drive}$ to $2^7 \cdot W_{drive}$, where $W_{drive}$ represents the width of a 1× driving transistor. As the number of synapses increases, most area of the network is consumed by the driving transistors. However, by using gated graphene to create analog spintronic weights, no transistor directly associated with the synapse is required. The only transistors in the network are the two in the dual-rail inverter in each cell. Assuming the two networks have the same charge-to-spin conversion efficiency, then the summed width of the synapse transistors in the digital-weighted spin-CNN would be the same as the transistor width of the inverter in the graphene spin-diffusive CNN. However, by merging the multiple synapse transistors into one wider transistor pair, significant footprint area could be saved: each cell saves at least $NoB \times N \times 6F^2$, where $N$ is the number of synapses per cell (Fig. 12) and the factor of 6 accounts for the gate contact pitch and transistor displacements. This calculation still underestimates improvement of graphene spin-diffusive CNNs by ignoring the two CMOS inverters (per cell) and the extra routing issues in the digital-weighted counterparts.

One challenge of the local gate programming in graphene spin synapses is the scalability of the graphene channel length. As shown in Fig. 7, the weighting range decreases with the shorter channel length, which is due to the increase of $\min(\lambda_s)/L$. This problem can be resolved by adding extrinsic SOC sources onto the graphene channel. For instance, as demonstrated by Yan *et al*. [49], gate tunable SOC on graphene can be introduced by placing $MoS_2$ on the top of graphene channel, which allows graphene spin synapses to be scaled down to shorter channel lengths.

It worth noting that the tunable graphene spin synapses could also be very attractive in general analog applications by providing low-energy and compact solutions for scaling and summation operations. If viewing the sensing MTJ as a spin-to-charge converter, one cell with multiple input graphene spin synapses could serve as a multi-input analog voltage adder/subtractor. In this case, STT-switching could potentially be eliminated, and so the energy consumption would be significantly improved, where only the signal-to-noise ratio during the spin injection, detection and transport would limit the critical current level for computation.

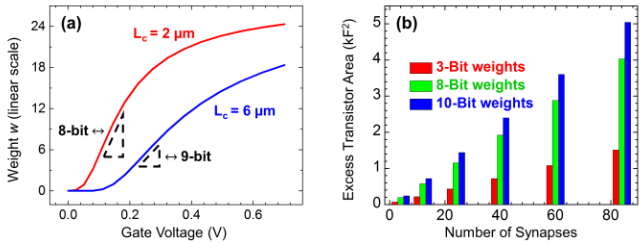

Fig. 12. (a) Gate voltage dependence of the applied weights on graphene spin-diffusive synapses (EY-dominated) at different channel length and the number of bits, *NoB*, required for binary weights. (b) Excess footprint area of digitally programmed spintronic CNN comparing to GrSD-CNN as the number of synapses increases.

## V. Conclusion

In this work, a spin-diffusive non-volatile CNN design using tunable graphene channels as the reprogrammable weights is presented. An efficient voltage-controlled spin-current source can be made using graphene due to its gate-tunable spin relaxation (Elliott-Yafet) and high sheet resistivity. Using a physics-based charge/spin circuit model, the device operation is verified using a low-pass filtering CNN as a case study. The energy consumption was found to be between 0.08-0.32 fJ/cell·synapse depending upon the computational accuracy. The proposed network also has both high fan-in and high fan-out properties as a result of its use of a mixed-mode operation whereby the spin domain is used to perform the weighted summation and the charge domain is used to perform the nonlinear transfer function. The transistor-free spin synapse also saves substantial footprint area compared to systems with digital weights particularly, where a greater number and more accurate connections are needed between cells.